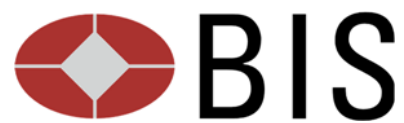

# BIS Working Papers
No 1040

## Quantifying the role of interest rates, the Dollar and Covid in oil prices

by Emanuel Kohlscheen

Monetary and Economic Department

September 2022

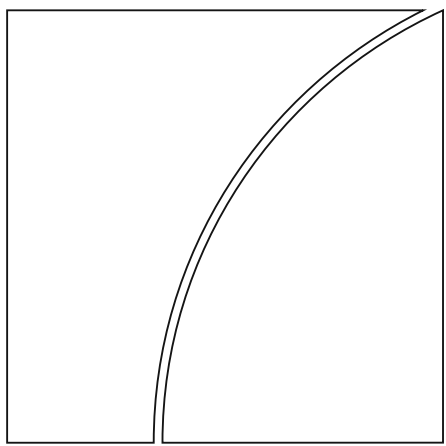

BIS Working Papers are written by members of the Monetary and Economic Department of the Bank for International Settlements, and from time to time by other economists, and are published by the Bank. The papers are on subjects of topical interest and are technical in character. The views expressed in them are those of their authors and not necessarily the views of the BIS.

This publication is available on the BIS website (www.bis.org).

# Quantifying the Role of Interest Rates, the Dollar and Covid in Oil Prices

Emanuel Kohlscheen[1,2]

## Abstract

This study analyses oil price movements through the lens of an agnostic random forest model, which is based on 1,000 regression trees. It shows that this highly disciplined, yet flexible computational model reduces in-sample root mean square errors (RMSEs) by 65% relative to a standard linear least square model that uses the same set of 11 explanatory factors. In forecasting exercises the RMSE reduction ranges between 51% and 68%, highlighting the relevance of non-linearities in oil markets. The results underscore the importance of incorporating financial factors into oil models: US interest rates, the dollar and the VIX together account for 39% of the models' RMSE reduction in the post-2010 sample, rising to 48% in the post-2020 sample. If Covid-19 is also considered as a risk factor, these shares become even larger.

---

[1] Senior Economist, Bank for International Settlements, Centralbahnplatz 2, 4002 Basel, Switzerland. *E-Mail address*: emanuel.kohlscheen@bis.org.
[2] I am grateful to Christiane Baumeister and to Deniz Igan for helpful comments and to Emese Kuruc for excellent research assistance. The views expressed in this paper are those of the author and do not necessarily reflect those of the Bank for International Settlements.

# 1. Introduction

Brent and WTI oil prices are ultimately asset prices. Their quotes have large reverberations for modern economies, not least for what concerns the evolution of headline inflation. Further, they may cause large swings in capital inflows and the national wealth of producing countries, as they ultimately lead to revaluations of the stock of assets that are still under the ground. As asset prices do, oil prices often react swiftly to relevant incoming news. Most recently, the WHO announcement of omicron as a Covid-19 variant of concern, on November 26th, 2021, highlighted this once more. Brent oil on that day fell from $81 to $71½ per barrel, a 12% reduction on impact.[3]

The asset price perspective suggests that incorporating financial factors could be important for understanding oil price movements, given that these factors vary continuously – with possible reverberations for oil markets. In view of this, this paper contributes to the literature by examining oil prices changes since 2010, disentangling in particular the effects of changes in the value of the dollar, market interest rates, overall market volatility and of Covid-19. The analysis is based on a flexible random forest model (in this case an ensemble of 1,000 regression trees). This method has the key advantage that it can unveil and easily accommodate eventual non-linear relations and interactions between variables.

The benchmark random forest model that is presented is easily replicable and provides a clearly superior description of relations. In-sample root mean square errors (RMSE) for monthly price variations are found to be 65% smaller than the AR(1) and OLS models which use the same set of 11 (daily frequency) explanatory factors. Indeed, even out-of-sample RMSEs are much smaller than the in-sample RMSEs of the econometric models. In proper 1- to 3-month ahead forecasting, the "1,000 regression tree model" delivers RMSE reductions that range between 51% and 68%, highlighting the relevance of non-linearities in oil markets.

Beyond comparing factor relevance and forecasting performance, the paper innovates in that it elicits the partial effects of individual variables on oil prices, discussing their economic interpretation. Overall, the findings underscore the importance of incorporating financial factors into oil models: US interest rates, the dollar and the VIX together account for 39% of the models' RMSE reduction in the post-2010 sample, rising to 48% in the post-2020 sample. If Covid-19 is also considered as a risk factor, these shares become even larger.

**Relation to the literature.**

The current article builds on many insights from the oil market literature. Reviewing these extensive contributions is beyond the scope of the present study. Instead, I highlight a limited selection of articles that are more directly related to this study.

The finding of strong relevance of current global activity levels for oil prices (here proxied by PMI activity surveys from 24 countries), is very much in line with the literature that aims to identify alternative proxies for global demand (e.g. Kilian (2009), Kilian and Zhou (2018) and Baumeister et al (2020) among others). The current paper clearly distinguishes between demand from advanced economies and emerging economies. The fact that emerging economy activity effects are relatively strong also aligns well with earlier findings by Aastveit et al (2015). This may be due not only to the growing shares of developing markets in the global economy, but also to their role as marginal buyers of oil.

---

[3] The dollar index varied very little on that day.

Salience of the clear inverse relation between the international value of the US dollar and oil prices is in accordance with earlier results in Akram (2009) and de Schryder and Peersman (2015), as is the positive association of oil with non-fuel commodity prices (Akram (2009) and Avalos (2014)). Further, the analysis finds an inverse relation with market risk, which accentuates when market risk increases sharply. The effects of the pandemic on oil markets is evident. More specifically, global Covid-19 deaths is found to have a highly asymmetric effect on oil prices. News about increasing casualties push the price of oil down, while news on further reductions in these counts produce little marginal effects.

Baumeister and Kilian (2014) evaluated econometric models for forecasting oil prices and concluded that using monthly data leads to more accurate forecasts than models based on quarterly data. This paper tries to take it one step further, by relying on daily data which can easily be updated on a continuous basis to produce better forecasts, also for horizons between 1- and 3-months. That said, as Baumeister (2016) put it "… *the price of oil will only be as predictable as its* (observed) *determinants, …*" (p. 154).[4]

Finally, the theory and use of machine learning techniques in economics is discussed in more detail in Varian (2014), Kleinberg et al (2015) and Mullainathan and Spiess (2017). Random forests were introduced by Breiman (2001), and have been shown to perform particularly well in economic forecasting (see for instance the comprehensive analysis of Fernandez-Delgado et al (2014), or the application to inflation forecasting in Medeiros et al (2021)). [5] [6]

**Outline.** The article proceeds as follows. Section 2 reports the sources of the key variable of interest (i.e. changes in spot crude oil prices), and the list of potential daily economic and financial explanatory variables. It then proceeds to present the machine learning model that underpins the analysis, and the marked reduction in prediction errors that it is able to deliver. Section 3 shows how much each factor contributes to this reduction in RMSEs and derives partial effects of the explanatory variables on oil prices. Section 4 analyses how the relative importance of key drivers has changed since the inception of the pandemic. Section 5 documents that the strong relative performance of the random forest model extends to pure forecasting exercises of oil price changes 1-, 2- and 3-months ahead. Section 6 presents robustness tests and Section 7 concludes.

---

[4] Observed added.

[5] Fernandez-Delgado et al (2014) compared the performance of 179 classifier models across 121 datasets, and found (the relatively simple) random forests to be the top performers.

[6] Within the asset pricing literature, Gu et al (2020) report large gains for investors that use machine learning. They attribute the predictive gains to "*allowing nonlinear predictor interactions missed by other methods.*"

## 2. Methodology

### 2.1 Data

The focus of the paper is to test how different economic and financial factors impact the price of oil. More specifically, it uncovers how the 22-day (monthly) log change of Brent oil price relates to the following list of potential explanatory variables: preceding month price changes (i.e. the log change between day *t-44* and *t-22*, as a "lagged dependent variable"); changes in the international value of the US dollar; in the interest rate in the US; in market volatility (CBOE VIX); in global Covid-19 mortality; in the US consumer price level (core PCE); in the price index of non-fuel commodities and in global economic activity.

US interest rates are the 2-year T-Bill yield, which has the advantage that it captures and is affected by both, conventional and unconventional monetary policy over the period under analysis (January 2010 – January 2022). Global economic activity is proxied by the Citibank Economic Surprise indices for advanced and for emerging market economies, as well as the average Markit's PMI indices computed for these two groups of countries. While the economic surprise indices are updated daily, based on whether incoming data on that day beat or underperform measured expectations, the PMIs are based on comprehensive surveys of corporates to measure current economic activity in each country. The complete list of variables that was used and their sources can be found in the Appendix.

All variables are in natural logs, and then converted to 22-day changes. The exception are interest rates and economic surprise indices (which are not in log). Overall, the sample contains 3,144 daily observations of the above variables. Note that even though the focus of the paper is on monthly variations, and not on very high frequencies, the methodology that is described below allows the use of all daily observations.[7]

### 2.2 The "1,000 Regression Trees Model"

To properly capture non-linear effects that may kick-in in oil markets – particularly for what concerns risk factors – the analysis is based on a methodology that can accommodate these well. More specifically, the benchmark model is based on random forests which were first introduced by Breiman (2001) (building on regression trees that were first presented in Breiman et al (1984)). These are non-parametric, yet highly disciplined models that are useful for uncovering non-linear and possibly complex interrelations between variables in a way that is by and large agnostic.

---

[7] This is because the random forest model classifies each observation individually, without reference to immediately preceding observations as in time series models. That said, the 22-day lagged change of the target variable is included among potential explanatory variables, so that the model can capture price momentum.

Random forests rely on an ensemble of regression trees. Put simply, regression trees aim to best predict the output variable (in this case the 22-day log change in Brent price). A tree is grown based on a computationally intensive algorithm that follows the sequence: i) randomly separate 1/3 of the observations for posterior performance testing (the test sample); ii) start with a single node containing all remaining 2/3 of the 3,144 observations of the sample (the "training sample"); iii) for all possible binary splits of all explanatory variables (features), compute the sum of mean square prediction errors that would result after the split; iv) split the dataset based on the variable and threshold level that delivers the lowest sum of MSEs; v) for each resulting node, return to step i). Simultaneously, at each split, only a random subset of the features is utilized. This process continues until a stopping criterion is met.[8]

An example of an estimated regression tree for the current application is shown in the Appendix, for a stopping criterion of $p$=1,000, where $p$ denotes the minimum number of observations per splitting node (the stopping criterion for the shallow tree). Essentially, outcome prediction is treated as a classification problem. Once the tree is fully grown, the algorithm evaluates in which terminal node a new observation would fall. It then predicts that the outcome will be equal to the average of the target variable for the "training set" observations in that node. The methodology is thus based on a purely data-driven method.

Robustness is achieved by basing outcome predictions not on one, but on a very large number of regression trees. That is the prediction is based on the average prediction of $N$ regression trees, which were grown based on $N$ different randomized subsamples of the dataset.[9] Figure 1 plots how the mean square errors for monthly Brent price declines very rapidly as the number of trees used in the random forest for monthly oil price variations grows.[10] [11]

---

[8] For a more detailed explanation see for instance the Statistics notes by Cosma Shalizi at www.stat.cmu.edu/~cshalizi/350-2006/lecture-10.pdf.

[9] Note that predicted values here, as in econometrics, refer to fitted values. These are based on contemporaneous explanatory variables. In Section 5, prediction refers to pure forecasting.

[10] Figure 1 plotted using $p$ = 10 (see later discussion).

[11] Mentch and Zhou (2020) find that "*the additional randomness injected into individual trees serves as a form of implicit regularization, making random forests an ideal model in low signal to noise (SNR) settings.*"

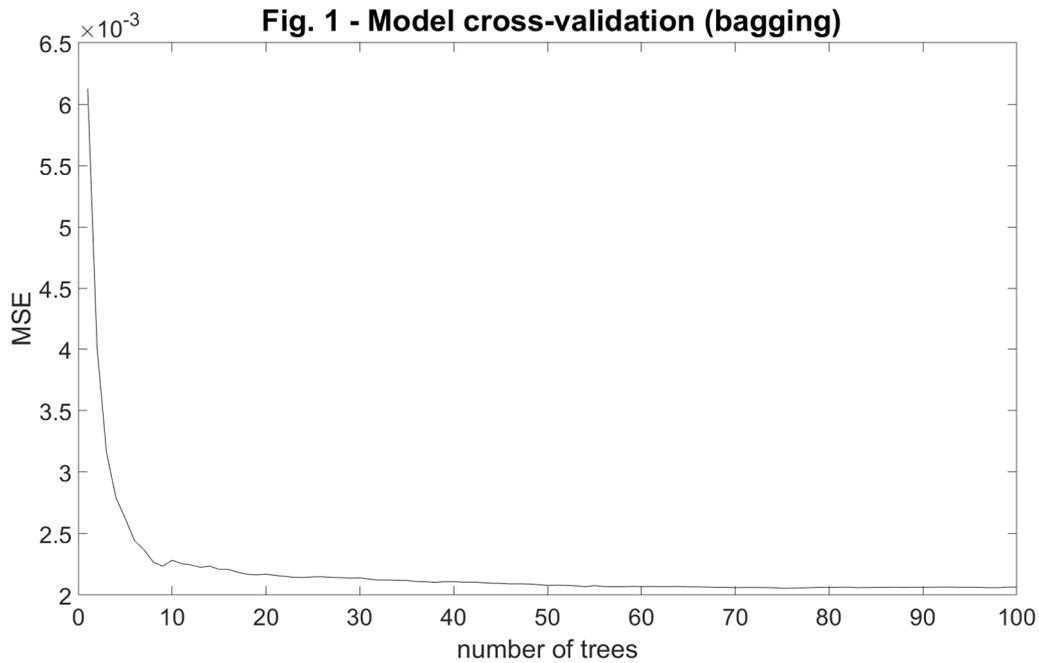

**Fig. 1 - Model cross-validation (bagging)**

The costs of overfitting within this framework are ultimately an empirical matter. In principle, deeper trees (i.e. lower values of $p$) could increase them and dampen out-of-sample prediction performance. Table 1 shows the RMSEs of the resulting models for alternative stopping criteria (first column) for random forests predicting Brent prices based on 100 and 1,000 regression trees. As expected, deeper trees (lower $p$ s) lead to smaller RMSEs in-sample (overfitting). Shallower trees (higher $p$ s) would tend to be more robust out-of-sample. Further, the prediction errors are typically somewhat smaller the larger the number of trees in the ensemble.

**Table 1 – RMSEs for Variation in Brent oil price vs. Minimum Size of Parent Splitting Nodes**

| min obs / splitting node | no. of reg. trees | root mean square errors (RMSEs) | | | | RMSE ratio (rel. OLS) |
| | | econometrics (in sample) | | ML | | |
| | | AR(1) | OLS | in sample | out of bag | |
|---|---|---|---|---|---|---|
| 4 | 100 | 0.1070 | 0.0745 | 0.0232 | 0.0433 | 0.312 |
| 5 | 100 | .. | .. | 0.0236 | 0.0427 | 0.317 |
| 6 | 100 | .. | .. | 0.0242 | 0.0430 | 0.325 |
| 8 | 100 | .. | .. | 0.0250 | 0.0437 | 0.336 |
| 10 | 100 | .. | .. | 0.0266 | 0.0448 | 0.356 |
| 20 | 100 | .. | .. | 0.0327 | 0.0477 | 0.439 |
| 30 | 100 | .. | .. | 0.0386 | 0.0517 | 0.518 |
| 40 | 100 | .. | .. | 0.0425 | 0.0543 | 0.571 |
| 4 | 1,000 | .. | .. | 0.0230 | 0.0421 | 0.308 |
| 5 | 1,000 | .. | .. | 0.0235 | 0.0424 | 0.315 |
| 6 | 1,000 | .. | .. | 0.0240 | 0.0424 | 0.323 |
| 8 | 1,000 | .. | .. | 0.0250 | 0.0430 | 0.336 |
| 10 | 1,000 | .. | .. | 0.0262 | 0.0436 | 0.351 |
| 20 | 1,000 | .. | .. | 0.0327 | 0.0476 | 0.440 |
| 30 | 1,000 | .. | .. | 0.0385 | 0.0511 | 0.517 |
| 40 | 1,000 | .. | .. | 0.0420 | 0.0532 | 0.564 |

Note: Sample period is from Jan 2010 to Jan 2022.

The benchmark random forest model is selected based on out of sample performance. Note that, overall, the "out of bag" RMSEs vary very little between $p$ values between 4 and 10, but then deteriorates much more rapidly as $p$ increases further. In fact, RMSEs are smallest for very deep regression trees ($p = 4$). What this points out is that the cost of overfitting in this application is essentially negligible.[12] Yet, to be on the conservative side, $p=10$ with 1,000 regression trees is selected as the benchmark model.

The last column of Table 1 shows that the 1,000 regression tree benchmark delivers a 64.9% reduction in RMSEs, relative to the OLS model that uses the exact same list of 11 explanatory variables. In other words, this confirms the relative strong performance of this widely used machine learning model.[13] [14]

---

[12] At least as long as p < 4 is excluded.

[13] This is in line with the conclusion of Medeiros et al (2021) for inflation prediction.

[14] The code is available upon request.

# 3. The baseline random forest model of Brent prices

## 3.1. Main drivers of oil prices (2010 − 2022)

The relative contribution of each factor for reducing the root mean square errors (RMSEs) of the Brent price model are shown in Figure 2. All 11 explanatory factors are found to be important drivers of oil price changes over the period, given their positive predictor importance statistics.[15]

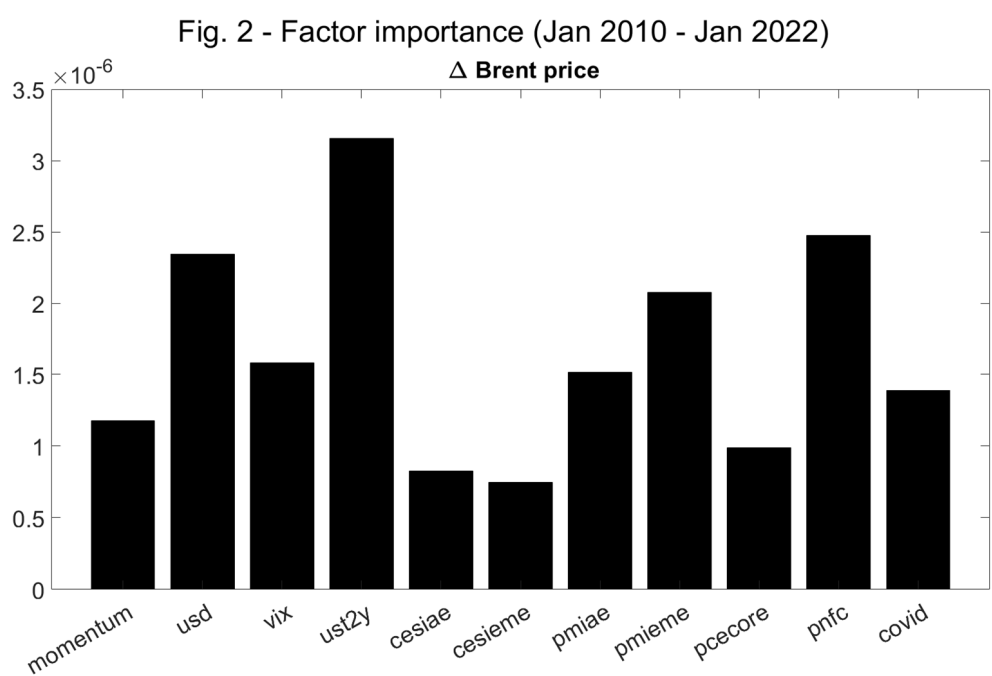

Fig. 2 - Factor importance (Jan 2010 - Jan 2022)

Interest rates in the United States (i.e. the change in the 2-year Treasury Bill has the largest imprint on prices among the variables), followed by changes in the non-fuel commodities price index (pnfc), the international value of the US dollar and activity levels in emerging markets, as surveyed by IHS Markit's PMIs (average of 15 countries).

When predictor importance factors are normalized, so that their sum equals 1, it becomes clear that financial factors − i.e. the dollar index, the 2-year T-Bill yield and the CBOE VIX volatility index − together contribute with no less than 38.8% to the reduction in RMSEs between January 2010 and January 2022. If global covid mortality is added as a risk factor, this share rises to 46.4%.

---

[15] Predictor importance factors are based on how much each factor contributed to reduce RMSEs after the splits in each tree. In other words, it is a metric on how many of the splits were based on that feature.

## 3.2. Partial effects

To examine how economic variables map into oil prices, we can analyze the partial effects of explanatory variables. This is done by varying the variable of interest, while all other variables are kept at their sample means. For each value of the variable of interest, the random forest then predicts an outcome (based on the average outcomes of such point over the 1,000 regression trees). This is essentially equivalent to coefficients of individual explanatory variables in an econometric model. The main difference here is that the random forest model is much more flexible, so that effects need not be linear (i.e. the slope coefficient is allowed to vary).

Figure 3 shows how the fitted oil price change varies with the change in global Covid-19 deaths. What is clear is that the Brent price declines when global casualty counts increase. In contrast, when they decrease (i.e. for negative values of Δcovid), the effect on the oil price flattens out (at a positive value).

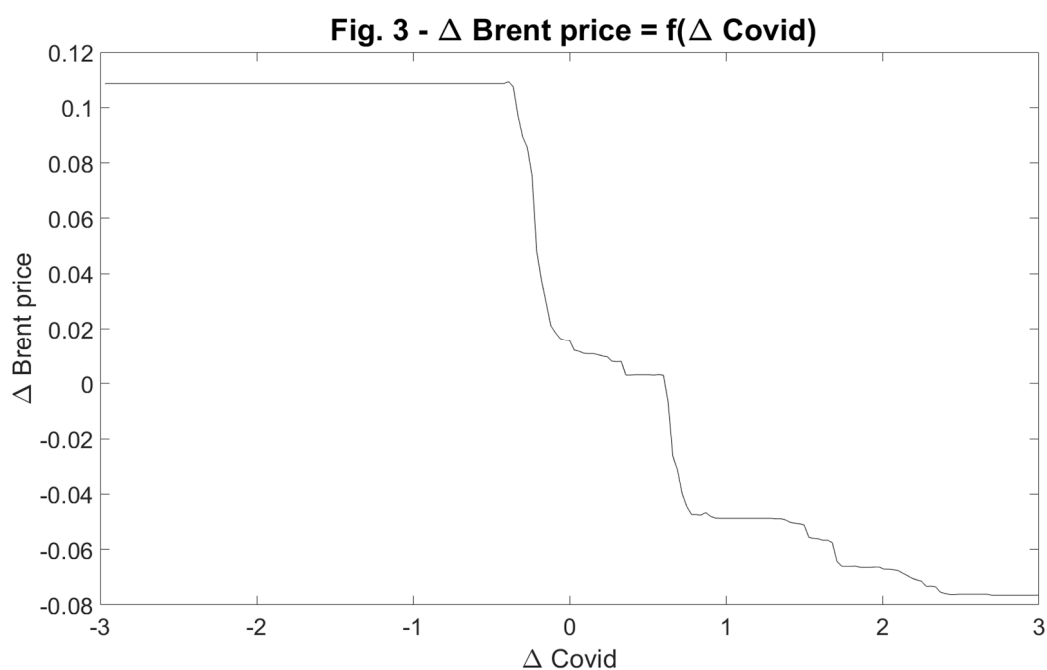

Similarly, Figure 4 shows how the partial effects of changes in market risk and in US Treasury Bill yields covary. Overall, strong increases in market risk (as proxied by the VIX), lead to an acceleration of Brent price drops. At the same time, for a given value of market risk, a lowering of bond yields tends to depress oil prices. This is likely a reflection of the fact that these may signal poorer growth prospects, and thus a more expansionary monetary policy (which is then reflected in a drop in the 2-year sovereign bond yields). Additionally, lower interest reduces the costs for leveraged oil producers, for instance in the shale oil sector.

**Fig. 4 - Δ Brent price = f(Δ VIX, Δ 2 yr yield US)**

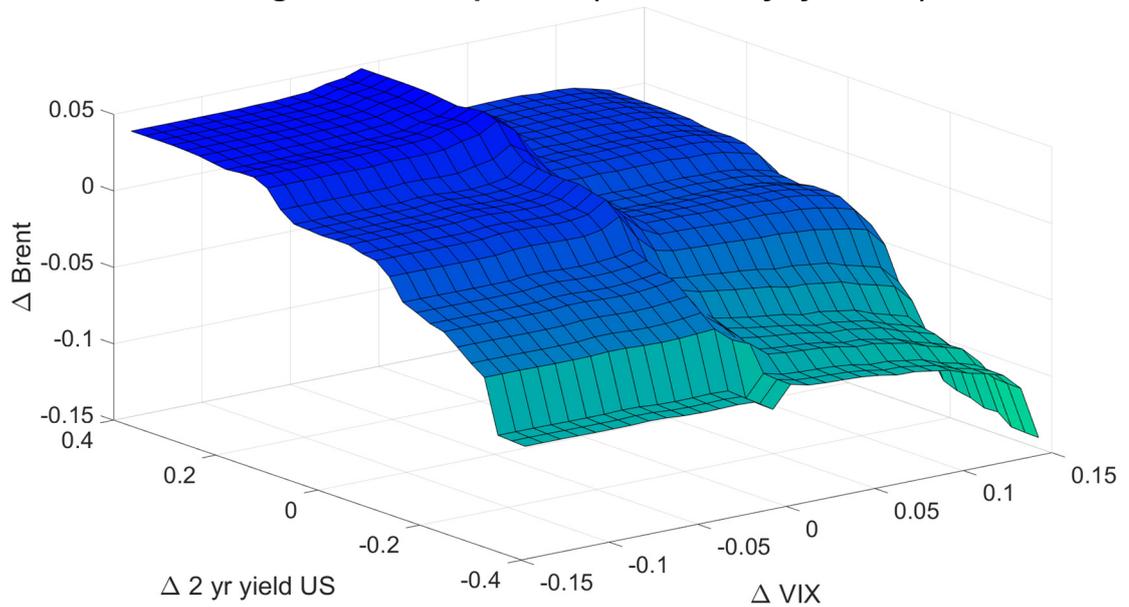

In turn, Figure 5 depicts how increases in other commodity prices spill over into higher oil prices. The positive association is in line with what has been documented earlier by Akram (2009) and Avalos (2014), among others. Importantly, the partial effects show a clear inverse relation of Brent prices with the international value of the dollar. This inverse relation is partly due to the fact that oil is invoiced in US dollars (the international numeraire). A strong dollar also means that the effective international purchasing power of other countries is reduced, depressing particularly the demand for oil that comes from emerging countries.

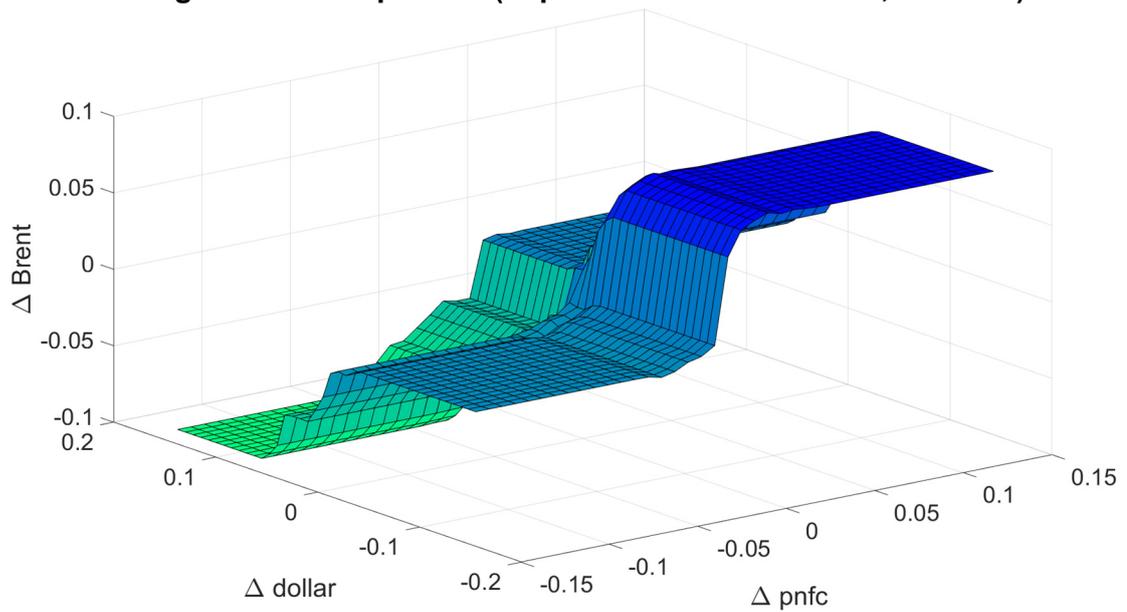

**Fig. 5 - △ Brent price = f(△ p non-fuel commodities, △ dollar)**

## 4. Post-2020 Sample

To examine variation across time, the exact same methodology was applied to a restricted sample, that starts in January 2020. This period is of course dominated by the Covid-19 pandemic. In this subsample, global pandemic related casualties are the third most relevant driver of oil prices (Figure 6).

The normalized factor importances by period, in Table 3, show that the relative importance of both, US Treasury bond yields and the dollar increased, while that of non-fuel commodities, PCE inflation and price momentum declined.

Overall, the joint relevance of the financial factors alluded to earlier increased from 35.5% in 2010-2019, to 48.3% post-2020. If Covid is included as an additional risk factor, from 35.5% to 65%.

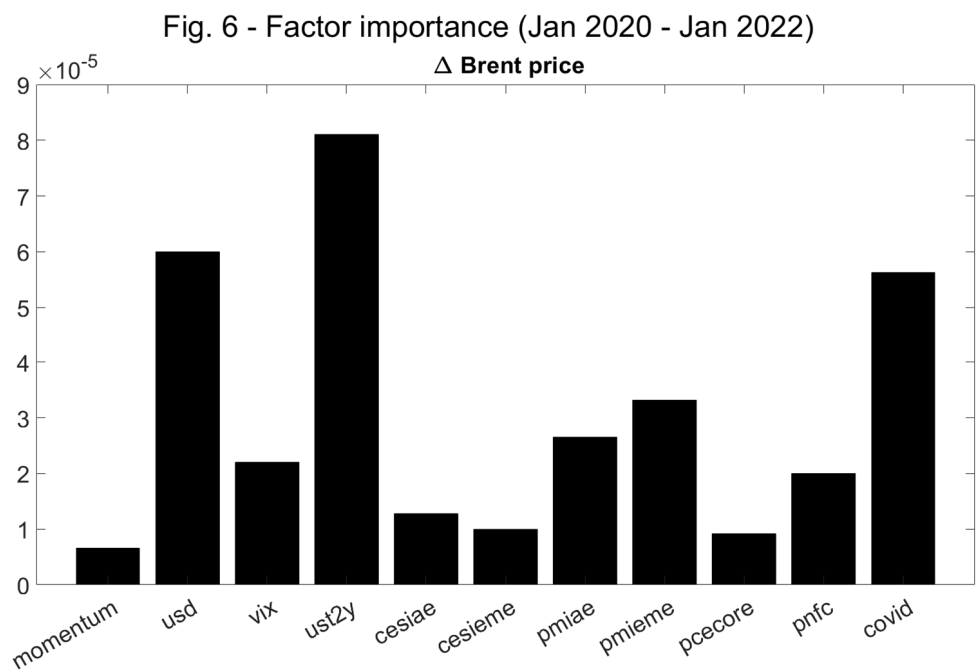

Fig. 6 - Factor importance (Jan 2020 - Jan 2022)

**Table 2 – Normalized factor importance by subsamples**

|  | Jan 2010 - Jan 2022 | Jan 2010 - Dec 2019 | Jan 2020 - Jan 2022 |
|---|---|---|---|
| US dollar | 0.128 | 0.155 | 0.178 |
| VIX | 0.086 | 0.094 | 0.065 |
| 2-year T-Bill rates (U.S.) | 0.173 | 0.105 | 0.240 |
| CESI AEs | 0.045 | 0.066 | 0.038 |
| CESI EMEs | 0.041 | 0.053 | 0.030 |
| PMI AEs | 0.083 | 0.080 | 0.078 |
| PMI EMEs | 0.114 | 0.085 | 0.098 |
| core PCE (U.S.) | 0.054 | 0.080 | 0.027 |
| price non-fuel commodities | 0.136 | 0.179 | 0.059 |
| global Covid-19 mortality | 0.076 | 0.000 | 0.167 |
| momentum | 0.064 | 0.102 | 0.020 |

Note: Normalized so that sum of all predictor importances equals 1.

## 5. Forecasting Future Oil Prices

The analysis in the previous section was based on prediction in terms of fitted outcome values. They reflected the large ability of the random forest model to explain or fit contemporaneous data better than standard linear econometric alternatives. This performance was confirmed by exercises in which the fit of data was tested on data that was not used in the computational construction of the respective regression trees.

The adaptation of the 1,000 regression trees benchmark to pure forecasting exercises is straightforward. It is done by changing the target variable from (log) changes in the price of oil between day $t-22$ and $t$, to changes between day $t$ and $t+22$, $t+44$ and $t+66$, while maintaining the same list of explanatory variables.[16] This is what is done in this section.

Table 3 reports the results of the above forecasting exercise. What is clear is that, irrespective of whether the 1-, 2- or 3-month ahead horizon is used, the random forest model outperforms the AR(1) and the OLS forecasting models by a wide margin. Relative to OLS, the reduction in RMSEs ranges between 50.6% and 68.2%.

**Table 3 – Relative Performance in Forecasting Oil Prices 1, 2 and 3 Months Ahead**

| | ML RMSEs | | ML RMSEs relative to | |
| | in sample | out of sample | out of sample | |
| forecast horizon | | | AR(1) | OLS |
|---|---|---|---|---|
| 1 month ahead | 0.030 | 0.048 | 0.448 | 0.494 |
| 2 months ahead | 0.028 | 0.052 | 0.325 | 0.340 |
| 3 months ahead | 0.034 | 0.059 | 0.297 | 0.318 |

Note: Ratios below indicate that ML outperforms AR(1) or OLS models. Based on model with 1,000 regression trees.

## 6. Robustness

The robustness of the general conclusions in Sections 3 and 5 to alternative specifications was checked and confirmed.

First, the benchmark model was re-estimated using alternative maximum tree depth parameters. Table A1 in the Appendix reports how similar results were obtained when the minimum number of observations per splitting node was alternatively set at 5, 6 and 8.

Second, Table A2 in the Appendix reports how similar results were obtained when the U.S. based WTI (West Texas Intermediate) price replaced the Brent as the target variable.

## 7. Conclusion

All in all, the analysis finds strong relative performance of random forests in predicting monthly oil price changes. The performance of this purely data driven method, relative to more traditional linear methods, is likely driven by its ability to accommodate nonlinear effects and interactions between explanatory variables.

---

[16] This follows the logic of the local projection method (Jordà (2005)).

The model which was presented relies on much more timely data inputs (at daily frequency). This is particularly important, as the model promptly incorporates changes that may be occurring in financial markets. The large role of these factors in reducing prediction errors only underscores the importance of interpreting spot crude oil as an asset price.

The complexity of the machine learning model deployed here was purposefully kept within limits, so as to enable the economic interpretation of the results.[17] Eliciting partial effects of variables proved to be useful in this respect. Future research could of course explore whether more complex models enhance forecasting performance further.

---

[17] This follows the Chakraborty and Joseph (2017) recommendation to limit complexity from the onset.

**Figure A1 – Example: Regression Tree (p = 1,000)**

**Table A1 – Varying max tree depth: normalized factor importance**

| | | min obs / splitting node | | |
| --- | --- | --- | --- | --- |
| | p = 10 (benchmark) | p = 5 | p = 6 | p = 8 |
| US dollar | 0.128 | 0.126 | 0.122 | 0.129 |
| VIX | 0.086 | 0.094 | 0.086 | 0.086 |
| 2-year T-Bill rates (U.S.) | 0.173 | 0.172 | 0.172 | 0.175 |
| CESI AEs | 0.045 | 0.046 | 0.047 | 0.046 |
| CESI EMEs | 0.041 | 0.042 | 0.043 | 0.042 |
| PMI AEs | 0.083 | 0.082 | 0.080 | 0.085 |
| PMI EMEs | 0.114 | 0.114 | 0.117 | 0.111 |
| core CPI (U.S.) | 0.054 | 0.055 | 0.055 | 0.055 |
| price non-fuel commodities | 0.136 | 0.134 | 0.138 | 0.136 |
| global Covid-19 mortality | 0.076 | 0.070 | 0.074 | 0.072 |
| momentum | 0.064 | 0.064 | 0.064 | 0.064 |

Note: Normalized so that sum of all predictor importances equals 1.

# Table A2 – WTI price: normalized factor importance

| | Jan 2010 - Jan 2022 | Jan 2020 - Jan 2022 |
| --- | --- | --- |
| US dollar | 0.117 | 0.134 |
| VIX | 0.075 | 0.046 |
| 2-year T-Bill rates (U.S.) | 0.109 | 0.126 |
| CESI AEs | 0.049 | 0.033 |
| CESI EMEs | 0.034 | 0.014 |
| PMI AEs | 0.117 | 0.105 |
| PMI EMEs | 0.151 | 0.154 |
| core CPI (U.S.) | 0.064 | 0.042 |
| price non-fuel commodities | 0.092 | 0.068 |
| global Covid-19 mortality | 0.134 | 0.240 |
| momentum | 0.057 | 0.038 |

Note: Normalized so that sum of all predictor importances equals 1.

**Supplementary Table – Summary statistics**

|  | mean | std. dev. | min | max |
|---|---|---|---|---|
| *all variables in 22-day changes* |  |  |  |  |
| Brent price (ln) | 0.0006 | 0.1086 | -0.8407 | 0.6235 |
| WTI price (ln) | 0.0002 | 0.1254 | -0.8653 | 1.2204 |
| 2 yr T-Bill yield (in %) | -0.0003 | 0.1412 | -1.1085 | 0.4255 |
| CESI AEs | 0.1073 | 24.81 | -143.50 | 138.70 |
| CESI EMEs | -0.1853 | 16.45 | -65.80 | 73.70 |
| USD NEER (ln) | 0.0014 | 0.0148 | -0.0432 | 0.0689 |
| VIX | -0.0092 | 7.9652 | -53.90 | 76.45 |
| PCE core (ln) | 0.0015 | 0.0011 | -0.0046 | 0.0062 |
| price of non-fuel commodities (ln) | 0.0022 | 0.0220 | -0.0622 | 0.0641 |
| global covid mortality (ln) | 0.0615 | 0.4655 | -0.4411 | 4.9738 |
| PMI AEs (ln) | 0.0010 | 0.0284 | -0.2644 | 0.1882 |
| PMI EMEs (ln) | -0.0003 | 0.0269 | -0.2175 | 0.1453 |

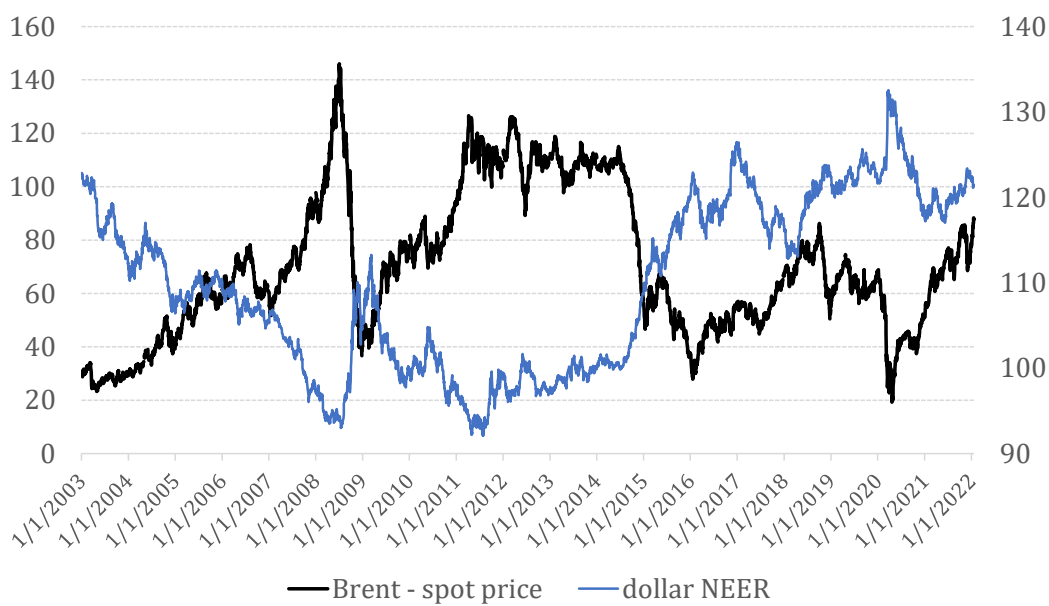

# Data Appendix

The database on which this study is based comprises 3,144 observations. All series are daily, with the exception of the core PCE index and PMIs, which are linearly interpolated to daily frequency.

List of variables and sources:

1. Bloomberg:
- Brent oil price in US $/barrel.
- West Texas Intermediate (WTI) oil price in US $/barrel.
- VIX: VIX volatility index traded on the CBOE.
- Interest rate: 2 year T-Bill yield, United States.
- CESI AE: Citibank economic surprise index, advanced economies.
- CESI EME: Citibank economic surprise index, emerging economies.

2. IHS Markit:
- PMI AE: GDP weighted average of the manufacturing PMIs of the United States, euro area, Japan, United Kingdom, Canada, Australia, Switzerland, Denmark and New Zealand. Interpolated to daily frequency.
- PMI EME: GDP weighted average of the manufacturing PMIs of China, India, Brazil, Russia, Mexico, Turkey, Indonesia, Korea, Malaysia, the Philippines, Thailand, Colombia, Czechia, Poland and Taiwan. Interpolated to daily frequency.

3. Our World in Data.org:
- Covid: daily number of global Covid-19 casualties, 7-day moving average.

4. Bank for International Settlements:
- Dollar index: nominal effective exchange rate, United States.

5. International Monetary Fund:
- pnfc: price index of non-fuel commodities.

6. Federal Reserve (FRED):
- Core PCE index: Core PCE index, interpolated to daily frequency.

The time period of the sample is 2 January 2010 – 20 January 2022.

## Previous volumes in this series

All volumes are available on our website www.bis.org.